\documentclass{pasj00}
\draft

\begin{document}
\SetRunningHead{M. Parthasarathy et al.}{Low Resolution Spectroscopy of 
Hot Post-AGB candidates}
\Received{}
\Accepted{}

\title{Low resolution spectroscopy of hot post-AGB candidates
II. LS, LSS, LSE stars and additional IRAS sources\thanks{Based on observations obtained at the Cerro Tololo Inter-American Observatory (CTIO), Chile.}}


%

\author{
Mudumba \textsc{Parthasarathy,}\altaffilmark{1,2,3}
John S. \textsc{Drilling,}\altaffilmark{4}\\
J. \textsc{Vijapurkar,}\altaffilmark{5}
and 
Yoichi \textsc{Takeda}\altaffilmark{1}
}

\altaffiltext{1}{National Astronomical Observatory of Japan (NAOJ) \\
2-21-1 Osawa, Mitaka, Tokyo 181-8588, Japan}
\email{m-partha@hotmail.com}
\altaffiltext{2}{McDonnell Center for the Space Sciences, Department of 
Physics, \\Washington University in St. Louis,
One Brookings Drive, St. Louis, MO 63130, USA}
\altaffiltext{3}{Aryabhatta Research Institute of Observational Sciences, \\
Nainital, 263129, India}
\altaffiltext{4}{Department of Physics and Astronomy, Louisiana State University,\\                           
Baton Rouge, Louisiana 70803, USA} 
\altaffiltext{5}{Homi Bhabha Center for Science Education, Tata Institute 
for Fundamental Research, \\V.N. Purav Marg, Mankhurd, Mumbai 400088, India} 

%


\maketitle

\begin{abstract}
Hot (OB) post-AGB stars are immediate progenitors of planetary nebulae (PNe).
Very few hot post-AGB stars are known. Detecting new hot post-AGB candidates
and follow-up multiwavelength studies will enable us to further understand the
processes during the post-AGB evolution that lead to the formation of PNe.
Case-Hamburg OB star surveys and their extension (LS, LSS, and LSE catalogues)
and IRAS (point source) catalogues are good sources for detecting new
hot post-AGB candidates from low resolution spectroscopy.
Spectral types are determined from low resolution optical spectra of 44 stars
selected from the LS, LSS, and LSE catalogues.                    
 Unlike the stars in the first paper,
the stars in this paper were selected using criteria other than positional coincidence with
an IRAS source with far IR (IRAS) colours similar to post-AGB supergiants and planetary nebulae. 
 These included high galactic latitude, spectral types of O, B, A  supergiants, 
emission lines in the spectrum and known spectral     
peculiarity. From the present study
 we find that 
 LSS 1179, LSS 1222, LSS 1256, LSS 1276, LSS 1341, LSS 1394,
 LSS 2241, LSS 2429, LSS 4560, LSE 16, LSE 31, LSE 42, and LSE 67 to be new hot post-AGB candidates.
Further study of these candidates is needed.
\end{abstract}

%

\section{Introduction}

    From an analysis of the IRAS point source catalogue data, several
post-asymptotic giant branch (post-AGB) candidates were detected 
\citep{partha86,partha89,pot88a,partha93b,kwok93,vanwin03}. 
These stars have far-infrared (IRAS)
colours similar to proto-planetary nebulae (PPNe) and planetary nebulae (PNe) \citep{ vander88,pot88b}.
Several of them are at high galactic
latitude, like HD 161796 (F3Ib) \citep{partha86} and LS II + 34 26 (B1Iae)
\citep{partha93b}. Multiwavelength studies of these objects by several investigators
have confirmed that these are indeed in post-AGB stage of evolution \citep{kwok93, vanwin03}.
The post-AGB stars 
seem to form an evolutionary sequence ( K, G, F, A to OB supergiant types)
in the transition region from the tip of the AGB to the early stages of 
planetary nebulae \citep{partha93b,partha93a}. 
 In order to understand the evolution from cooler to hotter post-AGB 
types and then into the young PN stage, it is important to detect and study
several hot post-AGB candidates.  
 \citet{vij97, vij98} and \citet{partha00}
presented the results of low resolution blue spectra of 40 IRAS sources with far-IR 
colours similar to PNe and PPNe \citep{vander88, pot88b}.
     A few of these objects have shown rapid changes in MK spectral
 type \citep{partha93a, partha95}.

The importance of the detection of new hot post-AGB candidates relies in objects
such as SAO 244567
(Hen 3-1357) and LS II +34 26. SAO 244567 evolved rapidly from  a B-type post-AGB star
into a young PN within a period of 20 years \citep{partha93a, partha95, bob98}. The variations
in the spectrum of SAO 244567 were dramatic. Another star LS II +34 26 initially classified as a massive
B supergiant \citep{turn84} turned out to be a rapidly evolving B-type post-AGB star
\citep{partha93a, smith94, garcia97}. Other such object are  LSE 162 (SAO 85766) 
 \citep{volk89, arkh07} (references therein) and LSIV -12 111 \citep{conlon93}.
 Another reason to detect more hot post-AGB
candidates is their abundance peculiarities \citep{mccau92} which are different
when compared with the chemical composition of cooler post-AGB stars \citep{vanwin03}.
Very few hot post-AGB stars are known and most of them are at high galactic
latitudes and several of them are not IRAS sources \citep{mccau92, moeh98}.
We started a program to identify new hot post-AGB candidates by obtaining spectra of
 selected 
stars from LS, LSS and LSE catalogues \citep{hard59,step71,dril94,dril95}, in the hope
that some of them would reveal their post-AGB nature by variations in the spectrum. And, if we
have a siginificant sample of hot post-AGB candidates subsequent chemical composition study
of these candidates may enable us to understand their abundance peculiarities.
LS, LSS and LSE catalogues are best sources
for selecting high latitude luminous hot stars. In this paper we present new spectral types 
for 44 of these stars based on observations made with the 1m telescope of the 
Cerro Tololo Inter-American Observatory (CTIO), Chile. Included are 
 several IRAS point sources which were not included in the first
paper \citep{partha00}.
\section{Selection criteria}

     The observing list consisted of stars selected from the LS, LSS,
and LSE catalogues (Table 1).
 The stars selected are either IRAS
point sources and or O, B and A supergiants at high galactic latitudes,
 according to MK types given in the 
literature. The sample is biased towards stars which show some post-AGB 
characteristic, i.e. positional coincidence with an IRAS point source with far-IR
colours similar to post-AGB stars and PNe \citep{vander88, pot88b},
 high galactic latitude or known spectral peculiarity, luminous O, B, and  A spectral types,
and emission lines in the spectrum.
Our sample includes 29 stars, which are not IRAS sources
so that we may be able to detect, post-AGB stars without dust shells,
similar to BD+ 39 4926 \citep{kod73}. We selected 38 stars from the LSS catalogue,
4 stars from LSE catalogue and two stars from LS catalogue. In the total selected sample of 44 stars only
 15 stars are IRAS sources (Table 2, Figure 1).

\section{Observations}

    Digital spectra of 44 selected (Table 1) southern candidates were
obtained during 19th to 27th April 1994 using the spectrograph and 2d-Frutti
two-dimensional photon counting detector on the 1m telescope
of the Cerro Tololo Inter-American Observatory in Chile. 
The exposure times ranged from 10 minutes to 30 minutes. Because
of limited observing time at our disposal we have observed
the selected stars only once.
The wavelength coverage is 3800\AA\ to 5000\AA, and judging 
from the comparison spectra, the resolution is about 3.5 \AA.
The data were extracted, wavelength calibrated, and normalized to 
the continuum with the standard IRAF software. Spectra of all the 44 stars
 are shown in Figure 2.
 Results of analysis of spectra
of 40 IRAS sources obtained during the above mentioned observing dates were given
in paper - I \citep{partha00}.

\section{Analysis}

\subsection{Spectral classification}
We have compared the spectra of our program stars with the spectra
of standard OB stars \citep{wal90}. Walborn and
Fitzpatrick made a digital atlas of the spectra of OB stars which
they observed with the same instrument, but at higher resolution than 
that described above. We found that smoothing the Walborn and Fitzpatrick
spectra by 3.5 \AA\ produced nearly identical looking spectra for the 
O9.5V star HD 37468, which we observed, and the Walborn and Fitzpatrick 
O9.5V standard HD 93027. The errors in the spectral
 atlas of Walborn  and Fitzpatrick are of the order of 0.2 to 0.3 subtypes.
Among the O-stars they could classify O9.5 and O9.7 and among B stars
they could classify B0.5 and B0.7. Spectral types for non OB stars were estimated
by comparison with the photographic atlas of \citet{yama78}.
 The spectral types determined from the
present investigation are given in Table 1. For non-OB stars we have also used
the digital spectral atlases of \citet{silva92}, \citet{jacoby84}, and \citet{pick98}, however these
atlases are of much lower resolution than our spectra.
The errors in our spectal types that we gave in Table 1 are of the order of 0.3 to 0.5 subtypes.
For example we were able to distinguish spectral differences among B1, B2,  and B3 stars.
The letters e, f, p, and n by the side of spectral types in Table 1 are of
standrard MKK notation.
Letter "e" indicates emission line(s) in the spectrum, "f" indicates O-type star
with emission lines, (f) indicates N III emission is present and the notation ((f)) signifies that in addition to strong He II 4686\AA~ weak N III
$\lambda$$\lambda$ 4634-4640-4642 emission is present \citep{wal90}, "p" indicates peculiar,
and "n" indicates broad lines. Spectra of all the stars listed in Table 1 are shown
in Figure 2 starting with star no. 1 in Table 1 (top fisrt column of Figure 2)
to star no. 44 (bottom of the second column of Figure 2). 
 Notes on some of the
objects is given in section 4.3.

\begin{table*}
\small
\centering
\caption{Spectral types of LS, LSS and LSE stars based on our spectra}
\begin{tabular}{|l|l|l|l|l|l|l|}
\hline
No.& Star & b & Sp. & New Sp.Type & m$_{V}$ & comment \\ 
\hline
 1&LSS 3169   & -4.24& W(C)          & pec.em.   & 13.2 &  PN [WC9]\\
 2&LSS 3299   & +3.99& WRh           & pec.em.   & 11.9 &  PN [WC11]\\
 3&LSS 207    & -4.40& OB+           & O6V((f))  & 10.9 &  Post-AGB ? \\
 4&LSS 3888   & -5.14& OB+           & O6V((f))e & 12.6 & PN  \\
 5&LSS 827    & +0.36& OB            & O6Vn      &  9.2 & O6:nne., in nebulosity\\
 6&LSS 3119   & +0.04& OB            & O8Iaf     &  9.2 &  HD117797(Oe); O8.5\\
 7&LSS 3418   & -9.66& OB:(ce),lep,h & O9Iae     & 11.0 & HD 141969; PN\\
 8&LSE 67     & -13.58&OB+           & O9IIe     &  12.2& -29 15495, PN, post-AGB\\
 9&LSS 1448   & -0.03& OB+r          & O9.5III   & 11.0 & CD -55 3196 \\
10&LSS 1947   & -0.51& OB            & O9.5V     & 10.1 & HD 305599 \\
11&LSS 4349   & +3.89& OB            & B0III     &  9.6 & -22 4400, Herbig Ae-Be\\
12&LSS 2354   & -1.55& OB-           & B0V       &  9.6 & HD 99898; B0.5V:, HII \\
13&LSS 1394   & -8.23& OB+ce,h       & B2:nep    & 10.5 & CPD -64 1154, Post-AGB? \\
14&LSS 2241   &-10.58& OB+           & B1Ib      & 10.1 & CD -71 730,  Post-AGB \\
15&LSE 42     &+14.59& OB+           & B1Ib      & 12.7 &  post-AGB \\
16&LSS 1245   & -6.57& OB            & B1III-V   & 11.4 & CD -56 2603 \\
17&LSS 1021   & -1.18& OBce,h        & B1II-Vne  &  9.1 &  HD 69425; B1Vpe \\
18&LSS 968    & +8.21& OB-           & B1V       & 10.7 & -17 2357\\
19&LSS 3434   & -0.46& OBh           & B1Vn      & 11.1 & -53 6867, Herbig Ae-Be\\
20&LSS 1256   & -6.28& OB+ce,le,h    & B2ne      & 12.2 &  Post-AGB? \\
21&LSS 1341   &-10.68& OB(ce)        & B2ne      &  9.6 & HDE 309784; Post-AGB \\
22&LSS 2429   & +7.51& OB+h          & B2:ne     & 12.7 & Flat continuum; Post-AGB? \\
23&LSS 1276   & +5.37& OB+h          & B2:nep    &  9.8 &  HD80834; B5nne, post-AGB? \\
24&LSS 866    & -5.94& OB-           & B2III     &  8.5 & -39 3775 = HD 65054\\
25&LSS 1263   & +6.72& OB-           & B2IIIn    & 10.1 & -38 5410\\
26&LSVI+5 5   & -2.09& OB            & B2IIIn    &  7.8 & +5 1279\\
27&LSS 1060   & -6.20& OB+           & B2III-V   & 12.7 & \\
28&LSS 1367   & +4.37& OB            & B2III-Ve  & 12.0 & CD -48 5103 \\
29&LSS 327    & -1.76& OB            & B2V       & 12.2 & \\
30&LSS 2832   & +0.79& OB+           & B2:V:     & 13.0 & \\ 
31&LSS 1339   & -7.34& OB-h          & B2Ve      & 10.7 & CPD -62 1290 \\
32&LSS 1392   & -5.96& OB+ce,le,h    & B2Ve      & 10.7 & HD 307467 \\
33&LSS 1996   & -9.91& OB            & B2Ve      & 11.6 & CPD -69 1417 \\
34&LSS 1213   & +1.90& OB-           & B2Vp      &  9.7 &  CD -42 4819 \\ 
35&LSS 1179   & -1.59& A1Ia:h        & B3Ibp     & 11.4 & CD -46 4657, post-AGB \\
36&LSE 3      &+12.26& OB+           & B3IIIe    & 11.5 & BD-18 4436, post-AGB\\
37&LSS 4560   & -7.37& OB            & B3IIIep   & 11.3 & Hen 3 - 1557, Post-AGB \\
38&LSS 1222   & -7.10& B7I-II        & B9Iap     & 11.6 & Post-AGB \\     
39&LSS 1340   & -3.19& A1II          & A1II      & 11.4 &PN, Binary Central Star \\
40&LSS 3309   & +8.65& A5Iab         & A3I       &  7.6 &  HD 133656, post-AGB \\
41&LSE 16     & +8.88& OB+           & A3I       & 12.0 & LSS 4079, post-AGB \\
42&LSVI+10 15 &  9.99& F5I           & F5Ia      &  8.1 & +10 1470, post-AGB\\
43&LSS 1033   & +7.49& OB:           & F7V       & 13.5 & \\
44&LSS 1120   & +7.22& OB+           & F7V       & 12.9 & \\
\hline
\end{tabular}
\end{table*}

\begin{table*}
\small
\centering
\caption{Stars which are IRAS sources}
\begin{tabular}{|l|l|l|l|l|l|}
\hline
Star & IRAS & 12$\mu$m (Jy) & 25$\mu$m (Jy) & 60$\mu$m (Jy) & 100$\mu$m (Jy) \\ 
\hline
LSS 207     & IRAS 07077-1825&  0.80 &   6.66&   0.40L& ----- \\
LSS 827     & IRAS 07502-2618&  8.49 &  69.32& 183.30L& ----- \\
LSS 1179    & IRAS 08487-4623&  0.25L&   0.91&   2.12 & ------\\
LSS 1340    & IRAS 09418-5703&  0.42 &   4.88&   6.40 & ----- \\
LSS 2354    & IRAS 11265-6239&  1.37L&  11.38&  28.98L& ----- \\
LSS 3169    & IRAS 13487-6608&  1.19 &   9.16&  11.75 & ----- \\
LSS 3299    & IRAS 14562-5406& 92.41 & 310.50& 176.60 & 71.30 \\
LSS 3309    & IRAS 15039-4806&  0.25L&   4.29&   3.61 & ----- \\
LSS 3418    & IRAS 15513-6600&  2.16 &  48.18&  43.81 & 19.58 \\
LSS 3434    & IRAS 15543-5342&  3.65 &  11.86&  26.55L& ----  \\
LSS 3888    & IRAS 16577-5018&  0.25L&   1.05&   1.99 &  ---  \\
LSS 4349    & IRAS 17408-2204&  1.61 &  13.28&  33.60 & 44.18 \\
LSS 4560    & IRAS 17591-3731&  0.35L&   1.22&   0.91 &   --- \\
LSE 3       & IRAS 17074-1845&  0.50 &  12.20&   5.66 &  3.47 \\
LSVI +10 15 & IRAS 07134+1005& 24.51 & 116.70&  50.13 & 18.72 \\

\hline
\end{tabular}
\end{table*}

\begin{figure}
  \begin{center}
    \FigureFile(180mm,160mm){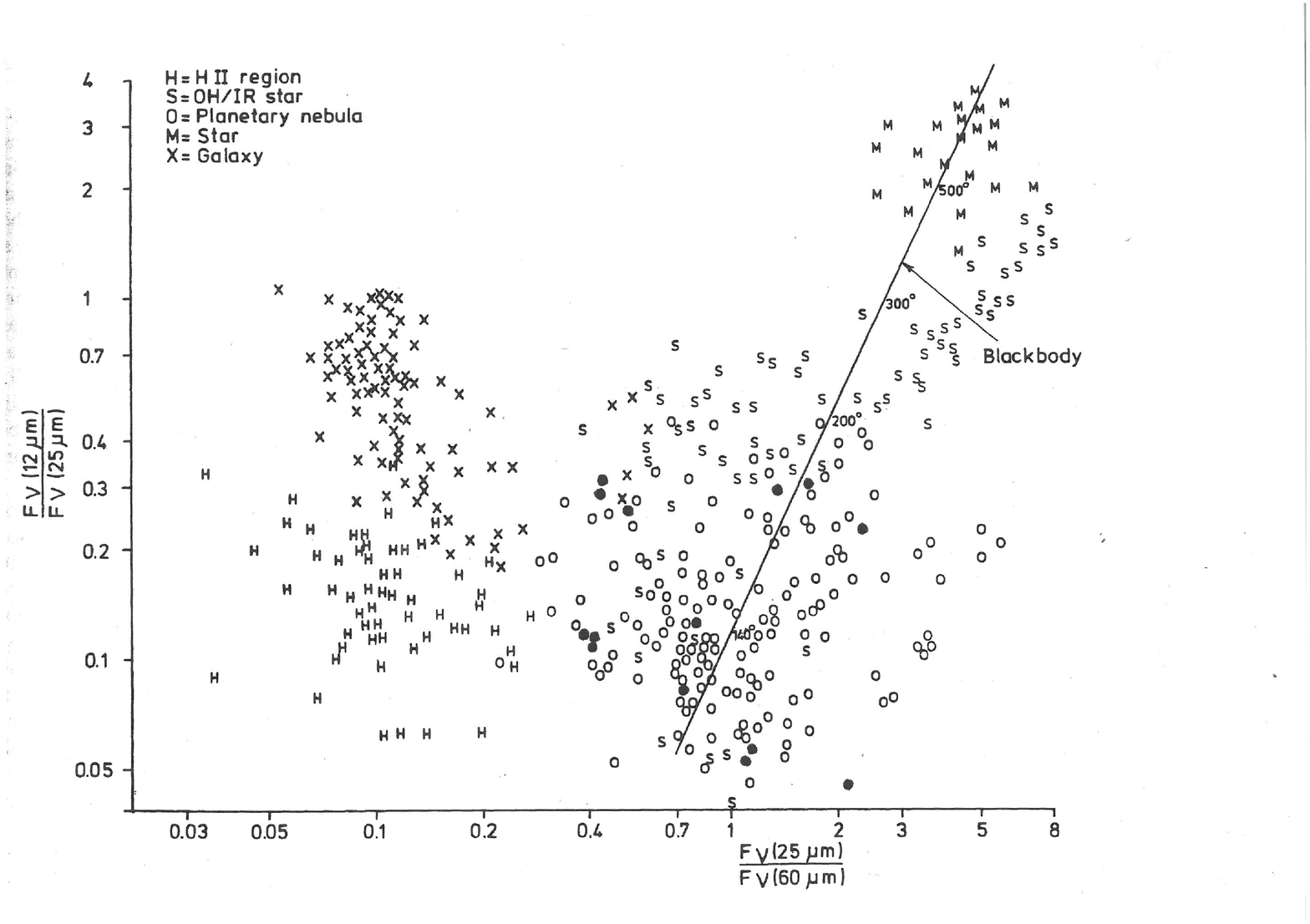}
  \end{center}
  \caption{IRAS colour-colour diagram adopted from the paper by \citet{pot88b}. The filled
circles are the objects listed in Table 2}
\end{figure}

\begin{figure}
  \begin{center}
    \FigureFile(150mm,180mm){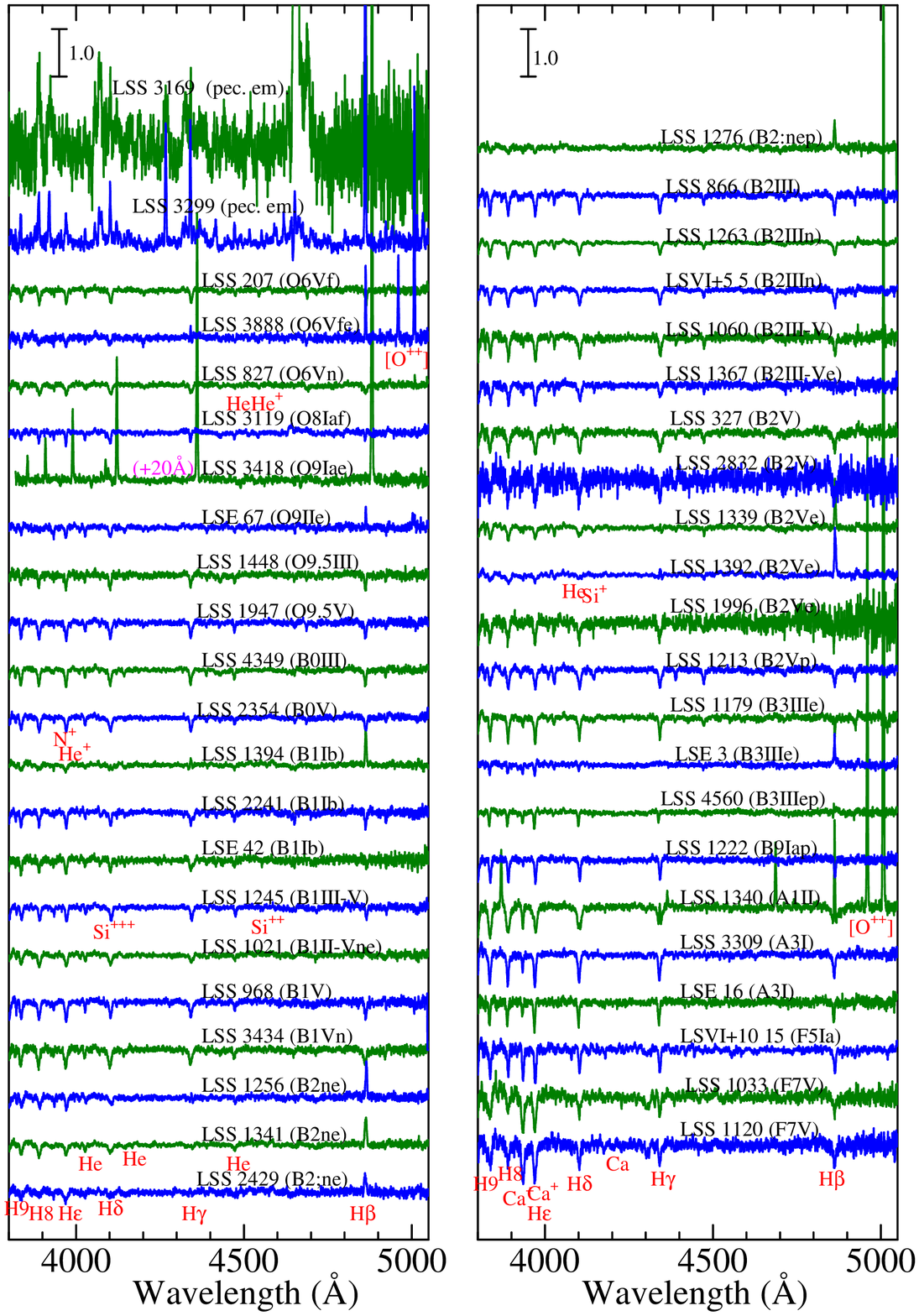}
  \end{center}
\caption{Specra of 44 stars (Table 1) (normalized with respect to
the continuum level), based on which we conducted spectral 
classifications. Each of the spectra are arranged according
the spectral type, and shifted by 1.0 relative to the
adjacent ones.
Positions of lines important for spectral classifications
(e.g., Balmer lines; He~{\sc i} 4026, 4121, 4144, 4471;
He~{\sc ii} 4009, 4551; N~{\sc ii} 3995; [O~{\sc iii}] 4959/5007;
Si~{\sc ii} 4128--30; Si~{\sc iii} 4552; Si~{\sc iv} 4089;
Ca~{\sc i} 4227; Ca~{\sc II} 3933/3968) are indicated in the figure.
Regarding LSS~3418 showing conspicuous emissions in Balmer lines, 
the wavelength scale is shifted by +20~$\rm\AA$ in order to 
avoid overlapping with the emission spectra of other stars. 
Note also the srong [O~{\sc iii}] emission in LSS~1340.
}
\end{figure}

\subsection{Location of the IRAS sources in the IRAS colour-colour diagram}

   In our sample of stars there are only 15 IRAS sources and they are listed
in Table 2. Their IRAS fluxes from SIMBAD are also listed in Table 2. The letter "L" on the side
of some of the fluxes indicates that  the error in the
flux value is  large. We have not listed the 100 micron flux in Table 2 for some of the objects
as their 100 micron flux is not relaiable (see SIMBAD data base). 
In Figure 1 we show the location of the objects (fillled circles)
listed in Table 2 in the IRAS colour-colour diagram \citep{pot88b}. The location of the star
LSS 207 is not shown in Figure 1 as its 60 micron flux quality is very low and  therefore the 25 to 60 micron
flux ratio is beyond the X-axis scale. The Figure 1 is adopted from the paper by \citep{pot88b}.
The 14 objects (Table 2) are in the region defined by PNe. PNe, PPNe and post-AGB 
supergiants have similar
IRAS colours \citep{pot88b} and occupy the same region in the colour-colour diagram defined by the
PNe (Figure 1). Most of the stars listed in Table 2 (Figure 1) are classified as post-AGB objects (Table 1).
 The evolutionary status of LSS 827, LSS 2354, and LSS 4349 is not clear. An occasional
H II region, or a nebula or a T Tau or Herbig Ae-Be star have IRAS colours similar
to PNe and post-AGB objects \citep{pot88b}.
 Notes on individual IRAS sources is given in the next section. 

\subsection{Notes on individual objects in Table 1}

\vskip .2cm
\# LSS 207 = IRAS 07077-1825
\vskip .2cm

 It may be a post-AGB star or a compact HII 
region. The far-IR (IRAS) colours are similar to PNe. It has a detached cold 
dust shell with a flux maximum at 25 microns. UBV photometric observations  
were made by \citet{dril91}. BVRIJHK photometric observations were made by \citet{fuji2002}.

\vskip .2cm
\# LSS 827 = IRAS 07502-2618 = CD -26 5115 = HD 64315
\vskip .2cm

It is listed as an emission line and variable star V402 Pup.
We find emission in the core of H${\beta}$. We also find emission lines at 4934\AA\ and 5009\AA\, which may be of circumstellar 
or nebular origin. In the Michigan spectral classification catalogue the spectral type 
of LSS 827 is given as A1(Ia)p. Spectral type variations appear to be
present. It may be an early type shell star. It is in the region of the
open cluster NGC 2467 and is involved in nebulosity. Cluster membership is 
not certain.

\vskip .2cm
\# LSS 1021
\vskip .2cm

 The H and HeI lines are filled in. It is listed as an emission line star
(MWC 855).

\vskip .2cm
\# LSS 1179 = IRAS 08487-4623
\vskip .2cm

  \citet{pre88} classified it as a new possible PN on the basis
of IRAS colours. In 1971 it was classified as a A1Ia star \citep{reed95}. However the
present spectrum is clearly shows that of B3Ibp star. The FeII lines are weak.
The HeI lines are broad with emission in the cores. The spectrum
indicates the presence of a shell. 
It may be a post-AGB star evolving rapidly to the early stages of a PN,
or it is an early type shell star.

\vskip .2cm
\# LSS 1222
\vskip .2cm

   It was earlier classified as B7I-II star \citep{reed95}.
   The spectrum suggests that it is a metal-poor B9Iap star. High
galactic latitude and a metal-poor supergiant type spectrum suggest that
it may be a post-AGB star similar to HR 4049 \citep{tak02}
and BD+39 4926 \citep{kod73}. It is not an IRAS source,
so is similar to the post-AGB
star BD+39 4926. Such stars can be termed as naked post-AGB stars.

\vskip .2cm
\# LSS 1245 = CD -56 2603
\vskip .2cm

    H and HeI lines are partly filled in by emission.

\vskip .2cm
\# LSS 1256
\vskip .2cm

   H$\beta$ and H$\gamma$ are in emission. The higher members of the Balmer
series also appear to be affected by emission.

\vskip .2cm
\# LSS 1276 = HD 80834
\vskip .2cm

       H$\beta$ and H$\gamma$ are in emission. Other higher members of the
Balmer series appear to be filled in by emission. It is classified
as a variable star QQ Vel.

\vskip .2cm
\# LSS 1339 = CPD -62 1290
\vskip .2cm

    Balmer lines up to H$\delta$ are in emission.

\vskip .2cm
\# LSS 1340 = IRAS 09418-5703 (PN G 279.6-03.1)(He2-36)
\vskip .2cm

     It is a PN central star. The central star is a binary with a A1II \citep{men78}
companion. He II 4686\AA~ is present indicating that the central star
is of very high temperature \citep{men78}. Our spectrum confirms the
presence of the A1II star companion to the central star.

\vskip .2cm
\# LSS 1341 = HD 309784
\vskip .2cm

   H$\beta$ and H$\gamma$ are in emission. HeI lines also appear to be
affected by emission. Many of the H and HeI lines seems to have
emission in the cores. High galatic latitude and the appearence of
the spectrum suggests that it may not be a massive pop. I OB star.
It may be a post-AGB star.

\vskip .2cm
\# LSS 1367 = CD -48 5103
\vskip .2cm

 H$\beta$ and H$\gamma$ lines appear to be affected by emission.

\vskip .2cm
\# LSS 1392 = HD 307467
\vskip .2cm

     Emission in Balmer lines.

\vskip .2cm
\# LSS 1394 = Hen3-356
\vskip .2cm

   Balmer lines in emission.

\vskip .2cm
\# LSS 1996
\vskip .2cm

    H$\beta$ and H$\gamma$ lines may be affected by emission.

\vskip .2cm
\# LSS 2241
\vskip .2cm

    We find it to be a high galactic latitude B1Ib star. It may be a
post-AGB star similar to LSII +34 26.

\vskip .2cm
\# LSS 2354 = IRAS 11265-6239 = V1087 Cen
\vskip .2cm

   H and He lines are affected by emission. They appear to be partially
filled in. The far-IR colours are similar to PNe. The presence of circumstellar
dust and B0V spectral type, and low galactic latitude indicates that
it may be a compact HII region.

\vskip .2cm
\# LSS 2429
\vskip .2cm

 H$\beta$ and H$\gamma$ are in emission. Other members of the Balmer series
also appear to be affected by emission. The spectrum is similar to that
of LSS 1394.

\vskip .2cm
\# LSS 3119 = HD 117797
\vskip .2cm

      The NIII lines are in emission. \citet{gomez1987} detected
strong carbon lines. They have estimated the mass-loss rate.

\vskip .2cm
\# LSS 3169 = IRAS 13487-6608 = He2-99
\vskip .2cm

        It is a planetary nebula. We 
classify the central star spectrum to be [WC9] which is in agreement
with the earlier classification \citep{ack92,ack03}.

\vskip .2cm
\# LSS 3299 = IRAS 14562-5406 = He3-1044
\vskip .2cm

    It is a planetary nebula.  We classify the central star spectrum to be
[WC11] which is in agreement with the earlier classification \citep{ack92, ack03}.

\vskip .2cm
\# LSS 3309 = HD 133656 = IRAS 15039-4806
\vskip .2cm
     Earlier spectral classification of this star was by \citet{vanwin97}
     It is an IRAS source with far-IR colours very similar to PNe.
 \citet{vanwin97} determined the chemical composition
from an analysis of high resolution spectra. They find it to be metal-poor.
We find the spectral type to be A3I. \citet{mon98}
analysed the UV (IUE) spectra of this star. They find E(B-V) = 0.32,
Teff = 8750K, log g = 2.5 and [Fe/H] = -1.0.  The A3I spectral type
derived by us is in agreement with the above parameters. High galactic
latitude, presence of circumstellar dust with colours similar to PNe
and underabundance of metals clearly suggest that it is a post-AGB star.

\vskip .2cm
\# LSS 3418 = IRAS 15513-6600 = He2-138 = PN G320.1-09.6
\vskip .2cm

      It is a planetary nebula \citep{ack92}. We
 classify the spectrum of the post-AGB
central star to be O9Iae.

\vskip .2cm
\# LSS 3434 = IRAS 15543-5342 = CPD -53 6867
\vskip .2cm

     The presence of circumstellar dust, low galactic latitude and
B1Vn spectral type suggest that it may be a Herbig Ae/Be type star.

\vskip .2cm
\# LSS 3888 = IRAS 16577-5018 = He2-187 = PN G 337.5-05.1
\vskip .2cm

  It is a planetary nebula \citep{ack92,koh01}.
 The spectrum of the central star is classified
for the first time. We find the spectral type of the central star to be
O6V((f))e.

\vskip .2cm
\# LSS 4349 = SAO 185668 = IRAS 17408-2204
\vskip .2cm

  \citet{mal98} classified it as a Herbig Ae/Be star. The IRAS
far-IR colours indicate a  cold circumstellar dust shell with high fluxes at
60 and 100 microns.

\vskip .2cm
\# LSS 4560 = IRAS 17591-3731 = HD 324802
\vskip .2cm

   The far-IR (IRAS) colours are similar to PNe.
    We find that H$\beta$ is in emission. H and HeI lines are partially
filled in. The far-IR colours, high galactic latitude and B3IIIep
spectral type indicate
that it is most likely a post-AGB star. UBV photometric observations
were made by \citet{dril91}.

\vskip .2cm

\# LSE 3 = IRAS 17074-1845 = Hen 3 - 1347
\vskip .2cm

     IRAS colours, high galatic latitude, and spectrum indicate
that it is most likely a hot post-AGB star \citep{gauba2003,uma2004}.

\vskip .2cm
\# LSE 16 = LSS 4079
\vskip .2cm

        A3I spectral type and high galactic latitude indicate that
it may be a post-AGB star. UBV photometric observations were obtained
by \citet{dril91}.

\vskip .2cm
\# LSE 42 
\vskip .2cm

     H$\beta$ is filled in. The B1Ib spectral type and high galactic latitude
indicate that it may be a post-AGB star. It may not be a massive pop. I B star.

\vskip .2cm
\# LSE 67  = CD -29 15495
\vskip .2cm
  It is a high galactic latitude OB+ star \citep{dril95}.
We find weak emission around [O III] 5007\AA~ 
 and H$\beta$ in emission. Based on these emission
lines, we conclude that it may be a new low excitation planetary nebula.
 The spectral type of the central star 
is found to be O9IIe.

\vskip .2cm
\# LS VI + 05 5 = HD 46106
\vskip .2cm

   Member (115) of cluster NGC 2244 \citep{ogura1981}

\vskip .2cm
\# LS VI +10 15 = IRAS 07134+1005 = HD 56126
\vskip .2cm

      Earlier spectral classification of this star is given in \citep{reed95}
     It is a well known and well studied post-AGB star. High resolution
spectra of this star was analysed by \citet{partha92, vanwin00, klo07}.
  It is metal-poor
([Fe/H] = -1.0) and shows over abundance of carbon and s-process elements.
It is a post-AGB star with 21-micron emission feature
 \citep{kwok93}.

\section{Discussion and Conclusions}

The presence of
circumstellar dust with far-IR colours similar to PNe, high galactic
latitude, OB supergiant type spectrum, and emission in the Balmer lines
are some of the characteristics of hot post-AGB stars. In the sample of observed
stars we found a few cases of B type dwarfs with H${\beta}$
emission. The presence of circumstellar material around Be dwarfs
indicates that they may be related to other Be stars, shell stars or
Herbig Ae/Be stars. In Table 1 we have a few other objects which are
not known 
to be post-AGB stars. They were included in the observing program as
some of them have IRAS colours overlapping with the IRAS colours
of some known post-AGB stars.  The comments given in Table 1 for some of the
stars are from the OB star catalogues and SIMBAD data base. For more
information on previous spectral classification, notes and comments
etc. refer to OB star catalogues and SIMBAD
data base. The m$_{V}$ given in Table 1 is from the SIMBAD data base.

    New spectral types for 44 O, B and A stars in the 
LS, LSS, and LSE catalogs have revealed several new hot (OBA supergiant)
 post-AGB candidates on the basis of either coincidence with IRAS point sources or high galactic
latitude. None of these objects shows any evidence of a changing spectral
type, but we plan to continue our monitoring program. 
Some of the post-AGB candidates we detected in our sample are not associated with
an IRAS source, indicating they do not have dust shells.
These are low mass objects and their evolution in the HR diagram
from the tip of the AGB is rather very slow and by the time they evolve
to A, B, O post-AGB spectral type, the dust shells seems to have disappeared
and they may never appear as planetary nebulae, hence they can be called 
naked post-AGB stars (e.g. BD +39 4926). Mutliwavelength study of post-AGB
candidates discussed in this paper is needed to further understand
their chemical composition and evolutionary stage. Some of the post-AGB candidates
given in Table 1 may show light variations similar to the high galactic latitude
hot post-AGB stars LS II +34 26 and SAO 85766 \citep{arkh07}.

  All the spectral data used in this study are given in the digital
form in electronic table E, which is available at the PASJ web site.

\bigskip
\bigskip
\section{Acknowledgements}

This research was supported in part by a grant from  
the National Science Foundation (NSF)
(AST-9819835) to JSD. MP is very thankful to Prof. Shoken Miyama
 and Prof. Ramanath Cowsik for their kind encouragement and support,
 JV participated       
in the work during her stay at IUCAA. We are thankful to Dr. Shashikiran Ganesh
for his help in converting the data into fits files. We are thankful to the
referee for helpful comments.

\end{document}